# A COMPARISON OF THE SPATIAL STATISTICS OF RANDOM AND DEFINED-SEQUENCE PHOTORESIST FILMS

**William D. Hinsberg[1]\* and Frances A. Houle[2]**

[1]Columbia Hill Technical Consulting, Fremont CA 94539

[2]Center for High Precision Patterning Science, Chemical Sciences Division and Molecular Biophysics and Integrated Bioimaging Division, Lawrence Berkeley National Laboratory, Berkeley CA 94720

\* Address all correspondence to William D. Hinsberg, columbia.hill@hinsberg.net

**Abstract**

*Background*: The resolution-line edge roughness-sensitivity tradeoff has motivated exploration of potential improvements using defined sequence polymers and polymer-bound photoacid generators and quenchers.

*Aim*: In this study we characterize the internal structures of positive tone photoresist polymer films formed from defined sequence polymers and compare them to random copolymers of the same composition. We model their imaging to connect initial to developable film structures.

*Approach*: We use a polymer packing algorithm to simulate films of diverse compositions and locations of photoacid generators and quenchers, using the composition of an ESCAP photoresist. We use a simple EUV exposure-deprotection algorithm to model developable image formation within them.

*Results*: In all cases, the spatial distribution of chemical moieties in the film for defined sequence polymers is nearly indistinguishable from random copolymers. We evaluate several exposure-deprotection scenarios, and find that a defined sequence copolymer has a distinctive developable image under certain circumstances.

*Conclusions*: The use of defined sequence polymers within a photoresist layer does not automatically result in improved imaging, however they do have some characteristics different from random polymers of the same composition. Further study of these characteristics may provide a route to improved control over the nanoscale imaging process.

**Introduction**

As the size of integrated circuit elements approaches the molecular scale, the photoresist films used to form those elements can no longer be approximated as a continuum material. This is particularly true for organic and hybrid chemically amplified resists exposed using EUV radiation. A few nanometer-scale printed features will contain only a small number of chemical moieties, including photoacid generator (PAG), base quencher (Q) and the base polymer. When combined with the small number of EUV photons typically employed to create the feature pattern in the film, these moieties lead to statistical variations of the chemical composition within the exposed and post-expose baked feature that in turn cause readily apparent deviations from the intended ideal shape. Two ways in which this statistical behavior is displayed are line-edge roughness (LER),[1] which measures how the edge of a printed feature deviates from an ideal straight line, and stochastic printing failures,[2] where a feature is missing from an extended array in a rare, random way.

Consideration of the factors that control image quality in chemically amplified (CA) DUV photoresists identified a resolution - line edge roughness – sensitivity (R-L-S) tradeoff.[3] Because of the statistical variations described above it is not obvious that how the tradeoff operates is the same in EUV lithography. The use of ionizing radiation EUV resists adds a significant level of complexity and additional source of inhomogeneity to the chemical composition of the exposed resist. The small number of EUV photons required to print a feature leads to a variation in the number of photons absorbed from one feature to the next. This variation, shot noise, follows Poisson statistics and leads to a significant difference in the extent of photolysis within a collection of features. The resulting formation of highly reactive species in the films such as positive and negative ions and free radicals is likely to be an additional factor that degrades resolution and LER, but is not necessarily directly connected to sensitivity in a simple way.

It is evident that improved control over resist composition variations and the attendant radiation chemistry may offer a means of reducing LER and improving resolution while maintaining a desired sensitivity. There are several aspects to consider for the current organic positive tone resist families. First, the components of an amorphous resist film are randomly located within the solid solution and their position will vary according to Poisson statistics. This aspect of the resist structure convolves with shot noise during EUV exposure. The extent of conversion of the matrix polymer to a soluble form depends on the kinetics of the acid catalyzed deprotection reaction which depends in turn on the details of the local radiation-induced chemical reactions, especially ones that generate acid. Second, aggregation of PAGs and Q compound the complexity of the chemistry. Third, the polymer characteristics add another dimension of statistical variability. For a typical resist polymer prepared by free radical polymerization of two or more monomer species, the final product will contain a statistical distribution of chain lengths and compositions. For positive tone systems, the solubilization of an individual polymer chain in aqueous base solution depends on the removal of hydrophobic groups by acid catalyzed chemistry until a threshold of the ionizable deprotection products is reached.[4] With chains of different length, composition and local acid catalysis conditions, each polymer molecule will have a different threshold where solubility is attained.

Several approaches to improve photoresist materials homogeneity have been pursued, for example the use of polymerization techniques that provide narrow molecular weight distributions,[5-7] the introduction of polymer-bound PAG and quencher,[8, 9] the use of molecular glasses,[10] and more recently defined-sequence polymers (DSP) with uniform lengths. Synthetic strategies to prepare DSP materials using methods such as reversible deactivation radical polymerization and solid-phase synthesis have been reviewed.[11] Preparation and applications of peptoid-based DSP materials to hierarchical nanomaterials for use in biomedical systems have also been described.[12] The potential for polypeptoids to improve control over the photoresist film for DUV and EUV exposures has recently been brought forward in studies describing their synthesis and imaging.[13, 14]

DSP-based resist materials offer a way to mitigate polymer-level imperfections through precise placement of resist moieties and PAGs. These structural characteristics can potentially provide further benefits within the amorphous photoresist films themselves if they influence the degree of randomness of photoresist components. A reduction in the random distribution of reactive moieties within the films may provide a way to control the details of inevitable radiation-induced chemical reactions occurring during EUV exposure. In this study, we investigate how imaging of a series of model CA resist films made of DSP materials compares to imaging of random copolymers of the same composition using a computational polymer packing model combined with a simple exposure-deprotection model. The simulations provide some initial steps toward an understanding of the characteristics of DSP-based CA photoresist films and the developable images they form within an amorphous layer.

**Materials and methods**

The goal of this study is to calibrate thinking about how controlled polymer sequence and length, resist component aggregation, and polymer-bound resist components affect the overall spatial homogeneity of an amorphous CA positive tone photoresist film, and thereby image quality. Similar considerations will apply to negative tone photoresists. Our primary metric for homogeneity is the spacing between the various monomers, so we employ a coarse-grained model of the resist film with a simplified representation of polymer chains as linear strings of spherical monomers. Monomers are connected by bonds with fixed bond lengths and bond angles, while the torsional angles along the chain are unconstrained. An ensemble of chains is randomly packed to form a dense amorphous glassy film. This representation of the polymers enables large film volumes to be calculated to improve precision of the statistical analysis of monomer distributions. The volume dimensions used in this work are 31 x 31 x 31 nm.

### A. Polymer sequences and film compositions

To evaluate the influence of polymer chain structure and location of PAG and Q on imaging, we examine several distinct types of polymer sequences within photoresist films of the same composition. Two are random copolymers where the PAG and Q are either bound or unbound to the photoresist polymer. The third is a series of DSPs with bound PAG and Q and one DSP with unbound PAG and Q. We also include nonpolymeric case which is a random distribution of monomers unconstrained by chain links.

The base CA photoresist film composition selected for evaluation is representative of a conventional positive tone EUV resist of the ESCAP-class.[15] It includes a copolymer of 4-hydroxystyrene (HOST), *tert*-butyl methacrylate(TBMA) and styrene(STYR). The composition also includes a PAG and a Q, with both either in polymer-bound form or as free molecules. PAG and Q in free molecule form may be random distributions of individual molecules or randomly sized aggregates with mean cluster size of 3 molecules. In all cases, the mole ratios are HOST:TBMA:STYR:PAG:Q = 50:30:6.7:10:3.3. **Table 1** summarizes the model compositions examined in this study.

**Table 1**. Combinations of polymer-PAG-quencher, with variations in sequence type, chain length and polydispersity.

| Identifier | Sequence[a] | PAG | Quencher | Chain length | Polydispersity |
|---|---|---|---|---|---|
| **R-BP-BQ-30-1.0** | Random | Bound | Bound | 30 | 1 |
| **R-BP-BQ-30-1.5** | Random | Bound | Bound | 30 | 1.5 |
| **R-BP-BQ-90-1.0** | Random | Bound | Bound | 90 | 1 |
| **R-BP-BQ-90-1.5** | Random | Bound | Bound | 90 | 1.5 |
| **R-FP-FQ-30-1.0** | Random | Free | Free | 30 | 1 |
| **R-FP-FQ-30-1.5** | Random | Free | Free | 30 | 1.5 |
| **R-FP-FQ-90-1.0** | Random | Free | Free | 90 | 1 |
| **R-FP-FQ-90-1.5** | Random | Free | Free | 90 | 1.5 |
| **R-AP-FQ-30-1.0** | Random | Aggreg. | Free | 30 | 1 |
| **R-AP-FQ-30-1.5** | Random | Aggreg. | Free | 30 | 1.5 |
| **R-AP-FQ-90-1.0** | Random | Aggreg. | Free | 90 | 1 |
| **R-AP-FQ-90-1.5** | Random | Aggreg. | Free | 90 | 1.5 |
| **R-FP-AQ-30-1.0** | Random | Free | Aggreg. | 30 | 1 |
| **R-FP-AQ-30-1.5** | Random | Free | Aggreg. | 30 | 1.5 |
| **R-FP-AQ-90-1.0** | Random | Free | Aggreg. | 90 | 1 |
| **R-FP-AQ-90-1.5** | Random | Free | Aggreg. | 90 | 1.5 |
| **R-BP-FQ-30-1.0** | Random | Bound | Free | 30 | 1 |
| **R-BP-FQ-30-1.5** | Random | Bound | Free | 30 | 1.5 |
| **R-BP-FQ-90-1.0** | Random | Bound | Free | 90 | 1 |
| **R-BP-FQ-90-1.5** | Random | Bound | Free | 90 | 1.5 |
| **R-FP-BQ-30-1.0** | Random | Free | Bound | 30 | 1 |
| **R-FP-BQ-30-1.5** | Random | Free | Bound | 30 | 1.5 |
| **R-FP-BQ-90-1.0** | Random | Free | Bound | 90 | 1 |
| **R-FP-BQ-90-1.5** | Random | Free | Bound | 90 | 1.5 |
| **S*x*-QP-BQ-30-1.0** | Sequence *x* | Bound | Bound | 30 | 1 |
| **S*x*-QP-BQ-30-1.5** | Sequence *x* | Bound | Bound | 30 | 1.5 |
| S*x*-QP-BQ-90-1.0 | Sequence *x* | Bound | Bound | 90 | 1 |
| S*x*-QP-BQ-90-1.5 | Sequence *x* | Bound | Bound | 90 | 1.5 |
| SF-FP-FQ-26-1.0 | Sequence F | Free | Free | 26 | 1 |
| SF-FP-FQ-26-1.5 | Sequence F | Free | Free | 26 | 1.5 |
| SF-FP-FQ-78-1.0 | Sequence F | Free | Free | 78 | 1 |
| SF-FP-FQ-78-1.5 | Sequence F | Free | Free | 78 | 1.5 |
| NonPoly | Monomeric | Free | Free | n/a | n/a |

[a] Defined sequences are listed in Table 2; here *x*=sequence A through sequence E

**Table 2** presents the DSP sequences utilized in this work. Each is a 30-mer with distinct placements of bound PAG and quencher. The 90-mers are 3 repetitions of the specified 30-mer sequence. The sequences are not exhaustive, but are designed to assess the effects of proximity on a chain. In each 30-mer, the first and third sets of 10 are identical, while the second, which contains a quencher but no styrene, differs from them. Sequence A serves as a reference, and results from it are the main ones reported here. Sequences B-E vary the position of the quencher relative to a PAG, and of the TBMA relative to the PAG in several combinations. It was found that in general, the statistical analysis results (see Computational Methods section) for polymer film structures using sequences B-E in the present work are nearly identical to those for sequence A, except for PAG-Q nearest neighbor distances.

**Table 2.** Defined Sequence Polymer Types

| Identifier | Repeating 30-mer Sequence[a] | Comments |
|---|---|---|
| A | HTH**P**STHHTH-HTH**P**H**Q**THTH-HTH**P**STHHTH | Nominal sequence |
| B | HH**T**P**STHHTH-HH**T**P**Q**THHTH-HH**T**P**STHHTH | Q at position 15 adjacent to PAG, each PAG has 1 adjacent TBMA |
| C | HTH**P**STHHTH-HTH**P**Q**THHTH-HTH**P**STHHTH | Q at position 15 adjacent to PAG, no TBMA adjacent to PAG |
| D | HH**T**P**STHHTH-HH**T**P**H**Q**THTH-HH**T**P**STHHTH | Q at position 16 away from PAG, each PAG has 1 adjacent TBMA |
| E | **T**HH**P**SH**T**H**T**H-**T**HH**P**HH**Q**TH**T**-**T**HH**P**SH**T**H**T**H | Q at position 17 distant from PAG, no TBMA adjacent to PAG |
| F | H**T**HS**T**HHH**T**H-H**T**HH**T**H**T**H-H**T**HS**T**HHH**T**H | Sequence A with PAG and Q removed |

[a] H=HOST, **T**=TBMA, S=STYR, **P**=PAG, **Q**=quencher

## B. Computational Methods

*i. Polymer packing and film formation - **Polyscope***

In this work, we introduce Polyscope, an open-source code[16] to construct films from polymer chains based on the polymer chain growth algorithm of Mueller et al (PolyGrow and Embed, Ph.D Thesis, Chapters 3 and 4).[17] PolyGrow treats chains as simple ball-and-stick assemblies, and builds each chain by choosing a random starting point in the film volume and adding monomers one by one with a randomly chosen torsional angle, while avoiding overlap with already-placed chains. For this, monomers are treated as spheres with a user-specified amount of allowable overlap during packing. As the film volume becomes densely packed, the procedure performs a local relaxation of chains near the chain end under construction to facilitate addition of more monomers.

Polyscope implements this chain growth and packing algorithm to form a representative photoresist film slice to which periodic boundary conditions are applied on all faces. Representative film slices constructed in this work are shown in **Figure 1**. Polyscope calculates the spatial distribution of polymer moieties within the film slice as a function of user-supplied key parameters, including polymer chain length, polydispersity, and random or defined sequencing. The code package enables detailed control over the film's chemical composition. An arbitrary number of different types of chemical building blocks can be specified. The building blocks can be monomer molecules of varying structures from which polymer chains are constructed and additive molecules such as PAG and Q. The sequence of building blocks in the polymer chains are specified to be random or a specific fixed sequence that is replicated in every chain. For random copolymers, all reactivity ratios (i.e. probability of a monomer attaching to a particular chain end building block) are assumed to be unity. PAG and Q can be specified to be part of the polymer chain (bound state) or dissolved in the photoresist film as an additive. Additives may be placed in the film as a random distribution of single molecules or aggregates of user-specified cluster size. Aside from aggregate formation, we do not include in the model other potential interactions between building blocks, such as hydrogen bonding or π- π stacking.

At present the roles such interactions might play in influencing the structure in spin-cast films are not well characterized and are outside the scope of the present study.

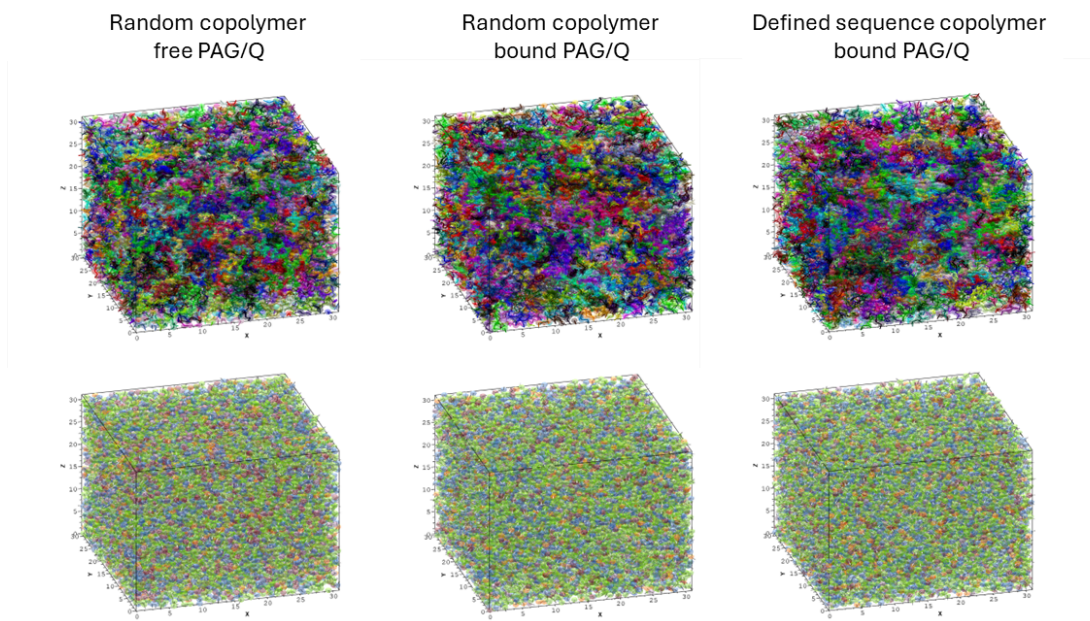

*Figure 1 – 3-D maps of 31nm x 31nm x 31nm film slices for selected typical compositions. The top row of slices shows placement of the polymer chains, colored to distinguish adjacent chains. The bottom row shows the locations of the monomers in the chains for the same slices; each monomer chemical type is colored differently*

A typical film slice simulation contains 30,000 building block units, each 1 nm in diameter.[18] These building block units are apportioned as monomers and additives of different chemical compositions according to the user's specifications. The packing density is 1 building block unit per cubic nm. The center-to-center distance of adjacent monomers in a chain is set to be 0.9 nm. The bond angle between adjacent monomers is 70.5°. During packing, a maximum 10% overlap of building block units is allowed, so the smallest possible distance between building block units is 0.9 nm.

After construction of a film slice is complete, PolyScope outputs a text file with the spatial location and chemical type of each monomer in each polymer chain. Each monomer is annotated with an index that marks which chain contains the monomer. Analogous location and chemical information for each nonpolymeric additive molecule are also contained in the text file. For a typical packing simulation (chain length =30 monomers, polydispersity =1.5) we calculate an average radius of gyration $R_g$ = 2.9 nm and an average end-to-end distance of 7.0 nm. These values are comparable to those obtained by Patsis *et al.* for molecular model calculations of a linear resist polymer of 30 monomers with monomer dimensions of 1 nm (2.7 and 6.3 nm respectively).[19, 20]

*ii. Statistical Analysis*

Statistical analyses of the simulation outputs are performed using the program *R,* an open-source software environment for statistical computing and graphics (version 4.3.2)[21] with the RStudio integrated development environment (version 2023.12.1.402).[22] The *spatstat* library for *R* (version 3.0-7) provides three-dimensional spatial statistical functions.[23] Additional statistical analyses (histograms, median values) are performed using the QtiPlot data analysis tool (version 1.2.1).[24]

A key metric of interest is the films' compositional uniformity. **Figure 1** does not reveal any visibly obvious differences, so to assess this we use the pair correlation function *g(r)*.[23] The function *g(r)* counts the average number of neighbors falling on the surface of a sphere of radius *r* for all chemical building blocks. This provides a measure of the probability of observing a chemical building block of a specific type separated by a distance *r* from a reference chemical building block, divided by the corresponding probability for a random Poisson spatial distribution. Hence *g(r)* is sometimes described as indicating "excess probability"; a *g(r)* value greater than 1.0 indicates clustering while a *g(r)* value less than 1.0 indicates a degree of regularity, or non-randomness.

    iii.    *Modeling film exposure and deprotection*

To examine the effects of EUV exposure on resist image spatial statistics using these slices, we apply a simple blanket exposure algorithm to the film compositions described above. For each composition, an energy dose is applied that in the end produces an overall partial deprotection of 50% of the TBMA protecting groups in the film slice. The 50% value was chosen to represent the extent of deprotection at the resist image line edge. This is where the film undergoes an abrupt change in dissolution rate and where stochastic effects are manifested.

The general exposure/deprotection protocol is as follows. To model EUV exposure, a photon absorption site is randomly selected in the film. A volume centered on the absorption site is defined, with a radius equal to the electron blur length $L$. It should be noted that this volume does not necessarily contain a single polymer chain, but parts of multiple chains. Within that volume there is a number of PAG molecules given by the PAG concentration. An average of $\varphi$ PAGs (where $\varphi$ is the quantum yield, acid molecules produced/photon absorbed) is randomly selected within the volume and converted to photogenerated acid. The absorption site and PAG selection steps are repeated until a specified number of acid molecules sufficient to produce 50% deprotection is produced.

Following the exposure process, a volume centered on a randomly selected photogenerated acid with a radius equal to the acid diffusion length $d$ is defined. Within this volume, all chemical building blocks are selected in random order and two types of reaction steps are allowed until no further reaction steps can occur. If the selected chemical building block is a TBMA with a protecting group, then the TBMA is converted to its deprotected product methacrylic acid (MAA). This step is allowed as long as there is TBMA remaining in the volume. If the selected chemical building block is a Q, then the acid is neutralized and no further deprotection can take place. The acid selection and reaction step processes are repeated until every photogenerated acid has been addressed.

For this study we use an electron blur $L$ value of 4.0 nm[25, 26] defining a volume which contains approximately 270 building blocks. We assume a $\varphi$ value of 4.0, that is each photon absorbed by a PAG produces on average 4 acids within the blur volume We examine two acid diffusion lengths for the deprotection step, with radius $d$ values of 2.0 and 5.0 nm.

**Results and Discussion**

The film slice compositions shown in **Tables 1** and **2** are analyzed to compare the spatial distributions of key components – TBMA with acid-labile protecting groups, PAG and Q – within a random and a defined-sequence polymer of the same composition. The simple exposure and deprotection steps are then used to evaluate the impact of these spatial distributions on formation of a developable image.

   A.  **Spatial Distribution of PAG**

One might anticipate that the spatial distribution of groups deprotected by acid would depend in part on the spatial distribution of PAG in the photoresist. **Figure 2** illustrates typical pair correlation functions *g(r)* for a selected range of different model compositions, where each composition contains identical numbers of the different chemical building blocks but in increasingly structured form. **Figure 2(a)** also includes, for reference, $g(r)$ obtained for a random point cloud distribution of discrete points (black dotted lines), showing a completely uniform distribution, as well as $g(r)$ for the case where the building blocks are present in the correct proportion but in non-polymerized form, the composition NonPoly in **Table 1** (dark blue curve). The random non-polymerized PAG-PAG $g(r)$ function shows low values inside a radius of 0.9-1 nm. This is a consequence of excluded volume – the centers of two building blocks cannot be closer than the sum of their radii (minus the allowed 10% overlap). Above 1 nm radius the curve overlays the point cloud curve, with some periodic structure that arises because the finite size of the building blocks excludes some of the center-to-center distances.

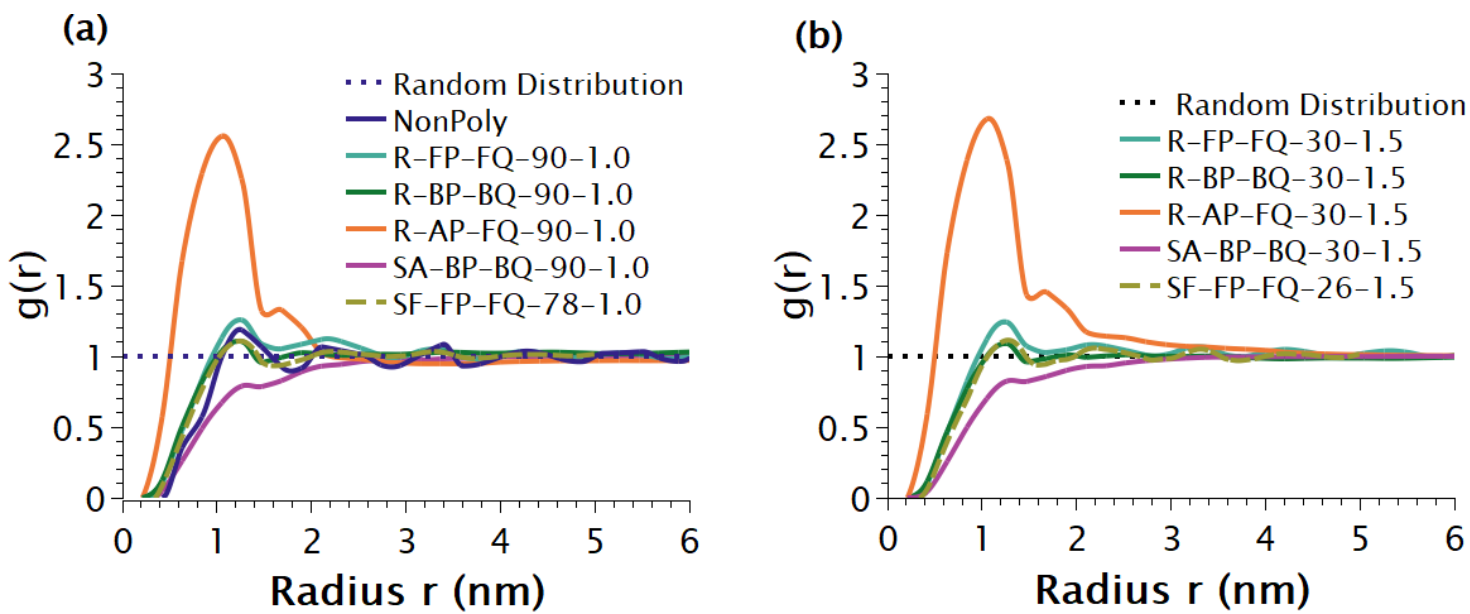

*Figure 2 – Pair correlation functions for PAGs in several film compositions. The defined sequences are sequences A and F in Table 2. (a) PAG-PAG correlations in a 90-mer chain of polydispersity 1.0, the navy curve is for a non-polymerized assembly of the same composition; (b) PAG-PAG correlations in a film with chains having a mean chain length of 30, and a polydispersity = 1.5. Note that although the two sets of curves are for different chain lengths and polydispersities, they are essentially identical.*

When the monomer building blocks (HOST, TBMA and STYR) are polymerized as a random copolymer, and the additive building blocks (PAG and Q) are present as free individual molecules (the composition **R-FP-FQ-90-1.0** in **Table 1**), the *g(r)* function (dark cyan curve) is little changed from the non-polymerized case. The orange curve depicts the *g(r)* function that results when the free PAG molecules are aggregated to a mean value of 3 molecules/cluster (composition **R-AP-FQ-90-1.0**). As expected, the probability that another PAG is a nearest neighbor (at a ~1 nm distance) is high compared to the unaggregated case. If the PAG and Q additives are polymer-bound, i.e. incorporated into the random copolymer (composition **R-BP-BQ-90-1.0**), the *g(r)* function (green curve) is very similar to both the nonpolymeric composition and the case where the additives are free individual molecules. These trends are in contrast to that for a polymer with defined sequence (composition **SA-BP-BQ-90-1.0** in **Table 1**), which produces the *g(r)* function shown as a violet line. The DSP pair correlation function at distances less than a 3 nm radius falls below that of a random distribution, indicating local order. All defined sequences with bound PAG and bound Q listed in **Table 2** yield essentially identical *g(r)* functions. However, a DSP composition where PAG and Q are free molecules (composition **SF-FP-FQ-90-1.0**), gives a *g(r)* function (dashed tan curve) equivalent to the random polymer cases.

**Figure 2(b)** shows *g(r)* functions for the same cases as **Figure 2(a)** with a mean chain length of 30 monomers and a polydispersity of 1.5. The PAG-PAG pair correlation functions in **Figures 2(a)** and **(b)** are quite similar to each other, indicating that chain length and polydispersity do not play a role in determining them.

It has been implied that PAG incorporation in a resist polymer improves the photoresist's spatial uniformity.[27, 28] Our calculations show that incorporation of the PAG functionality into a random polymer chain (a polymer-bound PAG) as described in **Table 1** does not intrinsically impart a more regular spatial distribution of PAG functionalities in the film. While binding PAG to a polymer likely reduces PAG aggregation/segregation compared to a conventional formulation that contains a poorly soluble free PAG,[29] our calculations indicate that there is no improvement to the uniformity of the PAG spatial distribution if it is polymer-bound compared to the case where the PAG is fully soluble (unaggregated) in the solid polymer.

In the DSP case, the *g(r)* traces in **Figure 2** show that the probability of one PAG having another PAG as a nearest neighbor is decreased compared to random polymers with free and bound PAGs. In the random case, all of a PAG's nearest neighbor sites have equal probability of containing another PAG. In the defined sequence case, however, the monomers vicinal to a PAG along the chain are strictly excluded by design from being another PAG, decreasing the overall likelihood of a PAG having another PAG as nearest neighbor.

Presumably, the ideal spatial distribution in the film of each of the resist components would be regularly spaced, which would minimize contributions of compositional inhomogeneity to resist stochastics. We can examine how closely the various compositions tested approach ideality by assessing the spacing between chemical building blocks of a given type. **Figure 3** shows histograms of PAG-PAG nearest neighbor distances for the cases in **Figure 2(a)**. **Figures 3(a)** and **(b)** show that for a random copolymer with bound and free PAG, the distance distributions are nearly identical. A large proportion of the PAG population (36% for bound and 35% for unbound) is within 1 nm of another PAG. The primary difference is that in the bound PAG case (**Figure 3(a)**) there is a spike at 1.46 nm due to geminal PAGs in the chain, that is, PAGs geminally bonded to the same building block. **Figure 3(c)** shows the histogram for the defined sequence case with free PAG and free Q. Its form is similar to the random polymer cases; 32% of the PAG population is within 1 nm of another PAG. **Figure 3(d)** shows the histogram for the defined sequence case with bound PAG and bound Q, which shows that there is still a significant proportion (20%) of PAGs within 1 nm of another PAG, smaller than the random copolymer cases. **Figure 3(e)** displays the histogram for the case where free PAG is aggregated within a random copolymer film; in this case essentially all PAG molecules (94%) have another PAG within 1 nm.

For comparison, **Figure 3(f)** shows histograms for near-ideal spatial distributions of PAG where regular, uniform-spacing is enforced. For the formulation used in this work (3000 PAGs distributed in a 30,000 $nm^3$ volume) we calculate that, precisely regularly spaced PAG molecules would be 2.15 nm apart. As shown in the coral curve of **Figure 3(f)**, when the positions of the PAG molecules are allowed to deviate randomly from the ideal uniform spacing coordinates by a small amount ($\sigma$ =0.025 nm), a nearest-neighbor distribution centered on the precise value is produced. In the purple curve in **Figure 3(f),** those positions are allowed to deviate randomly by a larger amount ($\sigma$ =0.24 nm), which not only broadens the nearest-neighbor distribution, but shifts it to smaller separation distances because of the significant proportion of PAGs that are closer to each other than in the uniform spacing case.

A comparison of **Figure 3(f)** to **Figures 3(a)-(d)** highlights the important finding from this work concerning the impact that chain packing alone has on the spatial distribution of resist components. The simulations show that the random chain packing that is characteristic of glassy amorphous polymer films leads to random, nonuniform spatial distributions of components, with a significant fraction of the component population located adjacent to another of the same component. This is found for both random sequence and defined sequence polymers, and indicates that it is highly unlikely that a way could be found to create polymers that can confer significant spatial uniformity within the amorphous structure.

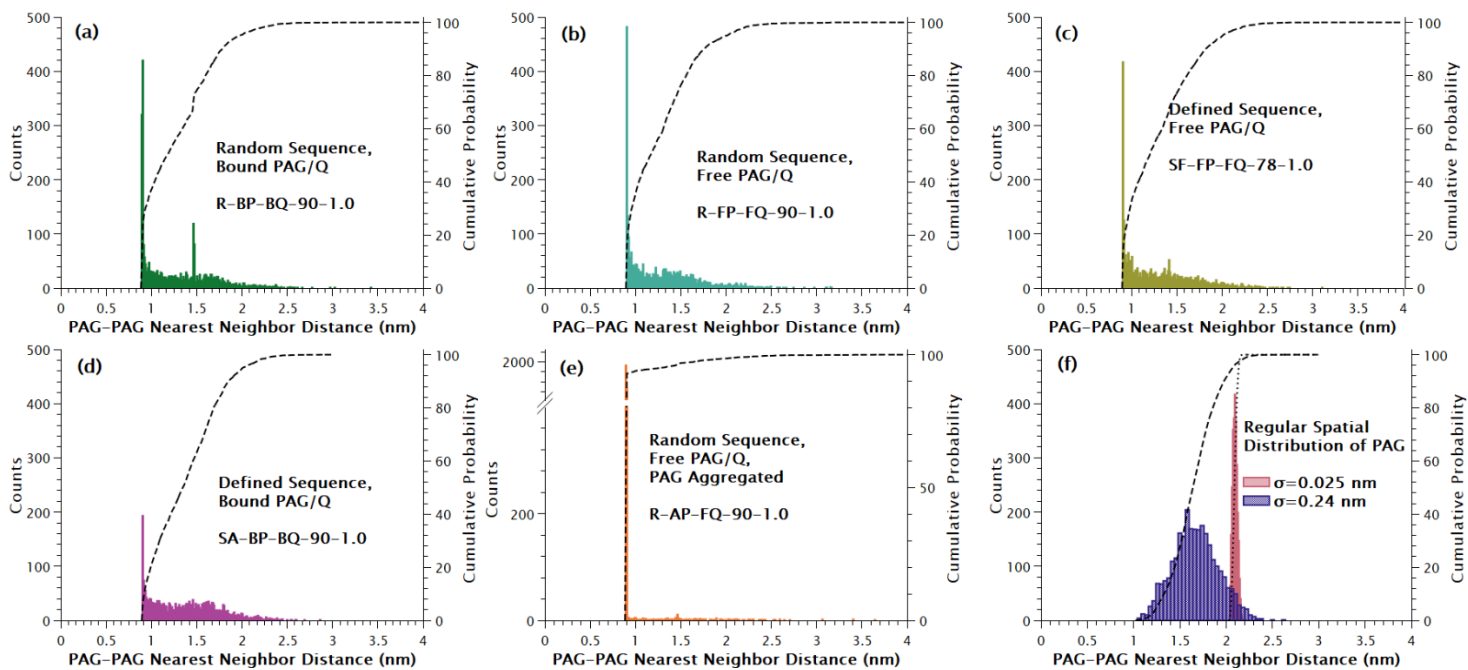

Figure 3 – Histograms of PAG-PAG nearest neighbor distributions for the cases shown in Figure 2(a), a 90-mer polymer with a polydispersity of 1.0, and 2 idealized cases. (a) Random polymer sequence with bound PAG and Q; (b) random polymer sequence with free PAG and Q; (c) Defined polymer sequence with free PAG and Q; (d) Defined polymer sequence with bound PAG and Q; (e) random polymer sequence with free PAG and Q where the PAG is aggregated; (f) comparison case of a regular spatial distributions of PAGs with random deviation in position of $\sigma=0.025$nm and $\sigma=0.24$nm. The dashed lines show cumulative probabilities.

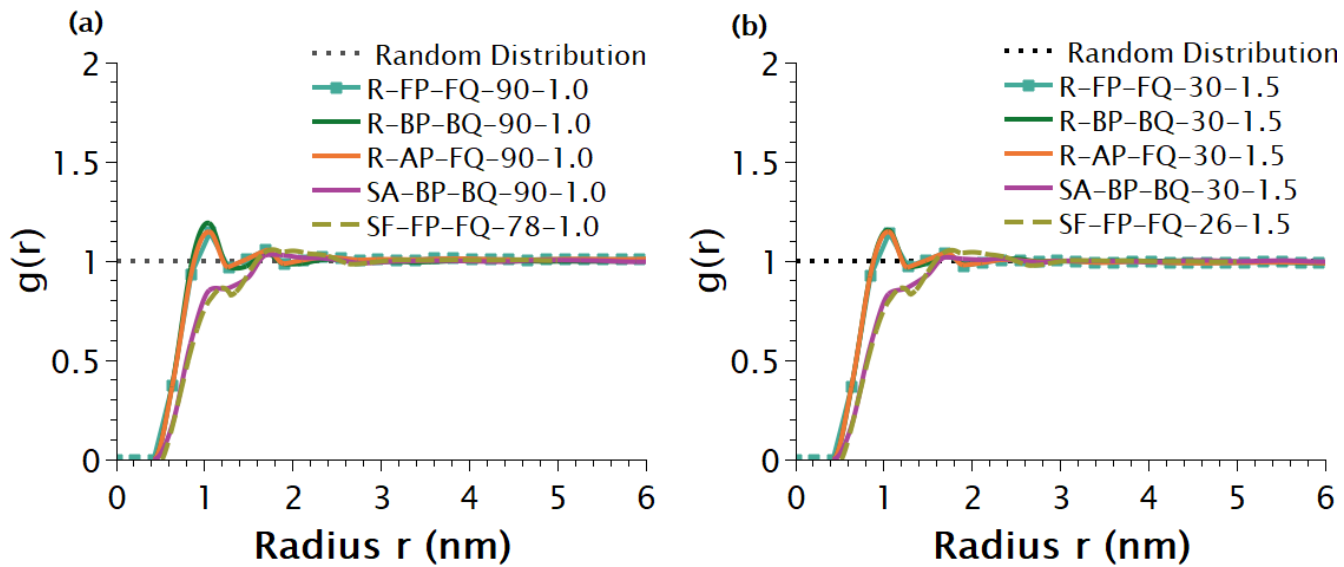

*Figure 4 – Pair correlation functions for TBMA in several film compositions. (a) TBMA-TBMA correlations in a 90-mer chain of polydispersity 1.0; (b) TBMA-TBMA correlations in a film with chains having a mean chain length of 30, and a polydispersity = 1.5. Note that although the two sets of curves are for different chain lengths and polydispersities, they are essentially identical.*

B. **Spatial distribution of deprotectable TBMA**

An analysis of the spatial distribution of TBMA, similar to that performed for the PAGs, leads to similar conclusions. **Figure 4(a)** illustrates *g(r)* pair correlation functions for the five model compositions examined in **Figure 2(a)**. All cases with random copolymers yield essentially identical *g(r)* functions that are consistent with a random spatial distribution of TBMA in the film volume. Both defined sequence cases deviate from the random cases in a manner analogous to that found for the bound PAG: the probability of one TBMA unit having another TBMA unit as a nearest neighbor is decreased compared to the other cases. All defined sequences in the present study (**Table 2**) were designed to avoid placement of TBMA monomers consecutively on the polymer chain, so this *g(r)* result is not surprising for the reasons discussed above for the PAG spatial distribution. The TBMA-TBMA pair correlation functions for cases where the mean chain length is 30 monomers with a polydispersity of 1.5, shown in **Figure 4(b)**, are virtually identical to those in **Figure 4(a)**, again demonstrating that chain length and polydispersity do not play a significant role in determining them.

**Figure 5** shows histograms of nearest TBMA-TBMA neighbor distance for the cases in **Figure 4(a)**. As found in **Figure 3** for the PAG distributions, a characteristic of a random spatial distribution is that a significant portion of the TBMA population has another TBMA immediately adjacent to it. In all cases where the polymer sequence is random, 74% of the TBMA has another TBMA within 1 nm distance. With the defined sequence cases, where TBMA is not immediately adjacent to another TBMA on the chain, this portion is reduced to 50% (free PAG/Q case) and 54% (bound PAG/Q case).

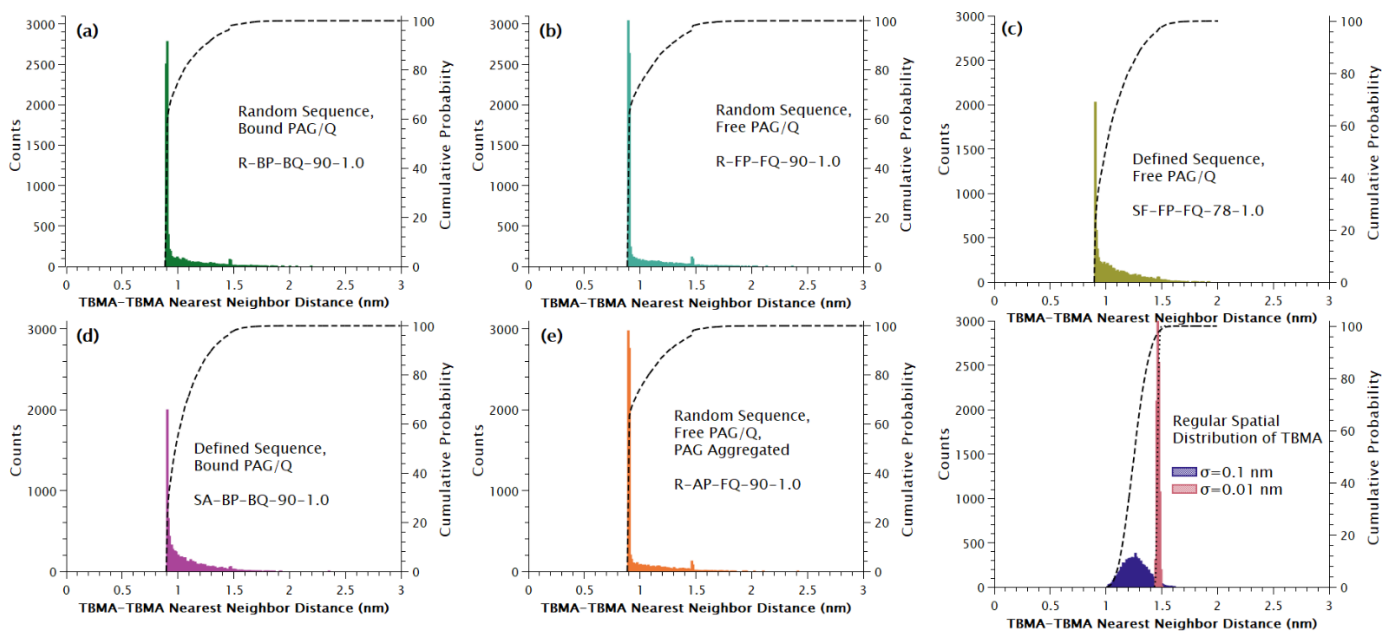

*Figure 5 – Histograms of TBMA-TBMA nearest neighbor distributions for the cases shown in Figure 4, a 90-mer polymer with a polydispersity of 1.0, and 2 idealized cases. Defined sequences are listed in Table 2. (a) Random polymer sequence with bound PAG and Q; (b) random polymer sequence with free PAG and Q; (c) defined polymer sequence F with free PAG and Q; (d) defined polymer sequence A with bound PAG and Q; (e) random polymer sequence with free PAG and Q where the PAG is aggregated; (f) comparison case of regular spatial distributions of TBMA with a random deviation in position $\sigma$=0.01nm and $\sigma$=0.1nm.* **The dashed lines show cumulative probabilities.**

## C. Spatial correlations of PAG-Quencher placements

The proximities of PAGs and Q in various film compositions may influence the imaging characteristics of the compositions, especially photoresist sensitivity and the clumpiness of deprotected polymer moieties. **Figure 6** shows histograms of nearest PAG-Q neighbor distances for several random and DSP structures, all for 90-mers with a polydispersity of 1.0. **Figures 6(a)** and **(d)** show that bound PAG/Q have a spike at the geminal distance of 1.5nm, as expected from **Figure 5(a)**, because the two species are held at specific distances by the chains. The distributions in **Figures 6(d)** and **(f)** show that if PAG and Q are separated by more than one monomer on a DSP, their spatial distribution becomes very similar to that of a DSP with unbound PAG/Q and all of the random polymer cases (**Figures 6(a)**, **(b)** and **(c)**). This indicates that a DSP with bound PAG and Q can only influence the PAG-Q spatial distributions if the PAG and Q are placed at a specific location close to one another on the polymer chain.

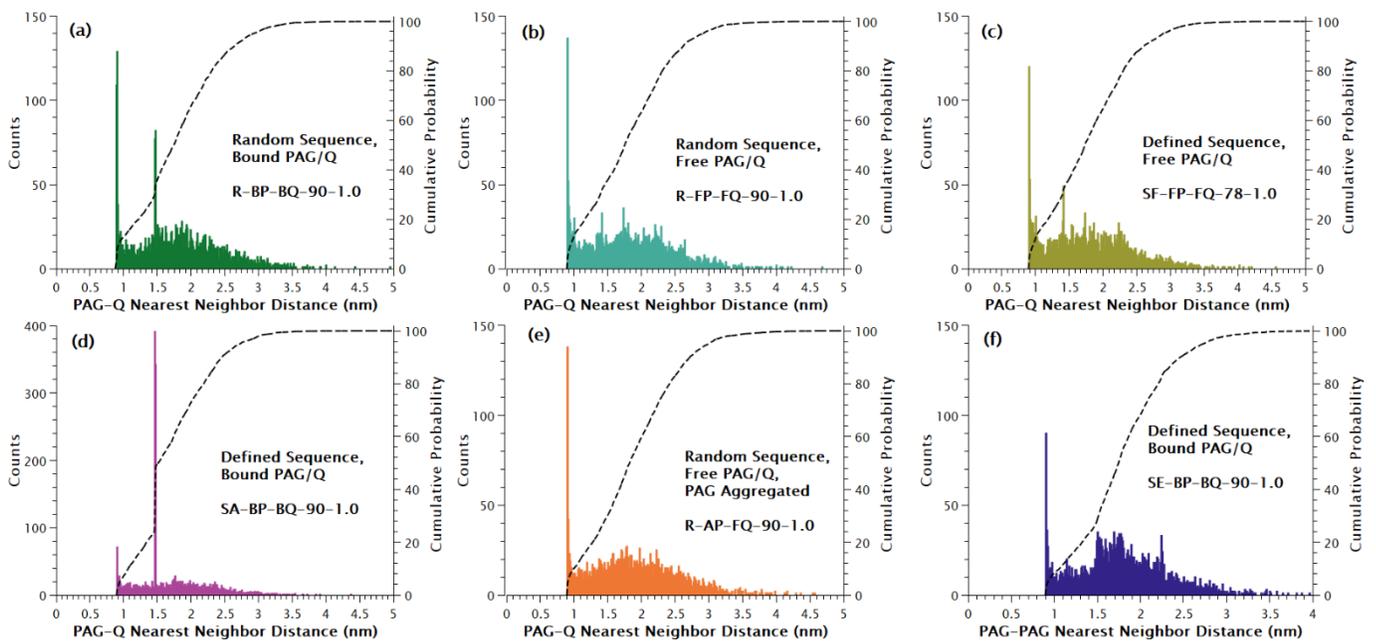

*Figure 6 – Histograms of PAG-Q nearest neighbor distributions for the cases shown in Figure 4, a 90-mer polymer with a polydispersity of 1.0. Defined sequences are listed in Table 2. (a) Random polymer sequence with bound PAG and Q; (b) random polymer sequence with free PAG and Q; (c) defined polymer sequence F with free PAG and Q; (d) defined polymer sequence A with bound PAG and Q; (e) random polymer sequence with free PAG and Q where the PAG is aggregated; (f) defined polymer sequence E with bound PAG and Q.*

### D. Impact of polymer film structure on exposure and deprotection

The EUV exposure and deprotection process substantially alters the spatial statistics of components within the resist film by changing their chemical identities. An individual photon absorption at a random location in the film is followed by production of a cascade of ionization events that produce electrons and diverse reactive intermediates localized near the absorption site, within a distance on the order of the electron blur length $L$.[27] In a positive tone photoresist, these products interact with resist components, ultimately forming some quantity of photogenerated acids near the original absorption site. Upon heating, these acids effect deprotection of the TBMA protecting group in a volume whose dimension is on the order of the acid diffusion length $d$. The net effect is the addition of an irregular "clumpy" compositional texture to the existing structure of the film, the spatial characteristics of which depend on these two lengths, the proximity of a quencher, the number of absorbed photons, and the efficiencies of the acid generation and deprotection processes.

For regions near the center of a resist pattern feature, where the exposure dose is relatively high and the initial acid distribution is dense, the deprotected volumes formed by those acids will overlap. Accordingly, those regions will be more uniformly deprotected and more likely to be fully soluble in the developer. For unexposed regions far from the nominal line edge, where the dose is near zero, any deprotected volume will be sparse, isolated and unlikely to dissolve because the number of ionizable groups has not reached the critical threshold (assumed to be 50% in this work). In the median dose range, near the line edge, the film's composition after deprotection transitions from insoluble to soluble. In this region, the irregular spatial composition is consequential and its non-uniform solubility will be manifested as edge roughness.

Of particular interest in this study are the spatial distributions of a deprotection product MAA, formed from TBMA. **Figure 7** compares *g(r)* functions for both acid diffusion length values of 2.0 and 5.0 nm. All random copolymers have a TBMA-TBMA correlation function *g(r)* consistent with random spatial distribution, with a decreased probability of adjacent TBMA groups in the defined sequence (see **Figure 4**). **Figure 7** shows that MAA produced from TBMA by the exposure and deprotection process leads to an MAA-MAA pair correlation function with values above 1 in all cases, consistent with the formation of clusters. This is to be expected since each acid can deprotect multiple TBMA until it is quenched or no TBMA remain within its diffusion volume. This is found for both random and defined sequence polymers. As shown in **Figure 7(a)**, when the acid diffusion length is short all cases with random copolymers show essentially identical *g(r)* curves. For the defined sequence cases, however, *g(r)* is in general lower than for the random polymer cases at MAA-MAA distances below 2 nm. For the longer diffusion length (**Figure 7(b)**), the pair correlations separate into two groups. The DSP and random sequence cases with free PAG and/or free Q have very similar MAA-MAA pair correlation functions above 2 nm radius, with the DSP case being more narrowly dispersed below that distance, as expected. The cases with bound PAG and Q, on the other hand, are more similar to those in Figure 7(a), indicating that acid diffusion length has less impact on the MAA-MAA nearest neighbor distances than when PAG and Q are unbound. The MAA distributions as a function of photoresist structure indicate that the spatial arrangement of resist components has somewhat less of an influence on the "clumpiness" of the deprotected regions if the acid diffusion length during post-exposure processing is short. If it is relatively long, on the other hand, it is possible that differences can be pronounced. This possibility is examined by considering the photoresist's solubility chain by chain.

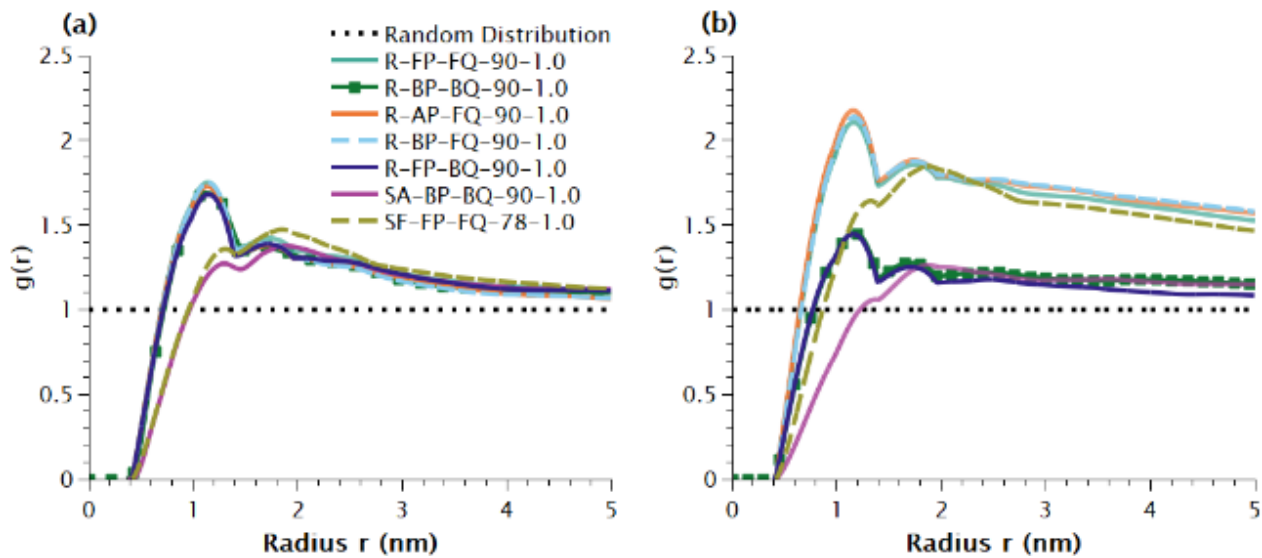

**Figure 7**. Pair correlation functions for MAA formed by exposure and deprotection as a function of radial distance in several film compositions with a 90-mer chain of polydispersity 1.0;. (a) MAA-MAA correlations in a film where the acid diffusion length is 2 nm; (b) MAA-MAA correlations in a film where the acid diffusion length is 5 nm.

    Resist polymer solubility in aqueous base is controlled by the fraction of ionizable groups on the polymer chain.[4] In all the compositions examined here, both HOST and MAA groups are ionizable. Ensembles of chains that comprise the random copolymer film slices include a broad range of different compositions with different fractions of ionizable HOST groups. Because the TBMA fractions have a similarly broad range across the chains, and exposure and deprotection events vary randomly, the ionizable MAA groups that form further alter each chain's solubility in complex ways. In contrast, the defined sequence polymer is a single molecular structure. When assembled into a film, is each DSP chain uniform in composition after exposure at a given local photon dose and deprotection? The pair correlation functions do not provide this information since they analyze only the overall packing within the film. If individual chains are uniform, one might anticipate that resist stochastic effects especially at the line edges would be attenuated.

We examine chain by chain statistics as follows. We analyze the composition of each ensemble of chains comprising a given resist composition both before and after exposure and deprotection. For an unexposed film, the properties of interest are the number of acid-labile TBMA groups and ionizable HOST groups in each chain. For an exposed film, each case, we find the distribution of numbers of MAA groups formed by deprotection of TBMA and the distribution of numbers of total ionizable groups (MAA+HOST) for each chain.

We have performed this analysis for all cases, and find that the results are well-represented by four particular cases: a random copolymer of polydispersity 1.0, with bound PAG and bound Q or unbound PAG and Q, and a defined sequence polymer of polydispersity 1.0, with bound PAG and bound Q or unbound PAG and Q. **Figure 8(a)-(c)** show results for the random copolymer case with unbound PAG and Q (composition **R-FP-FQ-30-1.0**). **Figure 8(a)** presents histograms of the numbers of TBMA and HOST in the chains prior to exposure. Both have very broad distributions, centered on the expected values of 0.346 for TBMA and 0.577 for HOST for a polymer of overall composition HOST:TBMA:STYR=50:30:6.7. **Figure 8(b)** shows histograms of chain compositions after exposure and deprotection for $d$ = 2 nm. Although the TBMA distribution is centered on 0.346 (probability normalized to 1), the resulting MAA distribution is centered on 0.167 at the applied dose reflecting the total extent of deprotection being limited to 50%. The final distribution of ionizable groups, MAA+HOST, which controls the chains' solubility, is approximately Gaussian and broadened compared to the initial HOST distribution. Only chains with MAA+HOST > 0.5 can dissolve. **Figure 8(c)** shows histograms of the chain compositions after exposure and deprotection for $d$ = 5 nm. With the longer diffusion length, there emerges a significant probability that there will be chains with no MAA. This result can be understood by considering that in our model a larger diffusion length will lead to a longer catalytic chain, that is, more deprotection events per acid formed, because the accessible film volume per acid-forming event is larger. Thus, fewer acids and fewer exposure events are required to meet an overall deprotection of 50% in the film. Fewer exposure events will lead to a more "clumpy" spatial character as indicated by the significant number of chains that contain no MAA (19-33% of the chains depending on the film composition).

**Figures 8(d)-(f)** presents the same analyses for random copolymer with bound PAG and Q (composition **R-BP-BQ-30-1.0**). The distributions are qualitatively the same as in **Figures 8(a)-(c)**, with shifts in all the distributions toward smaller median values (e.g., TBMA centered at 0.30 and HOST centered at 0.5, the values expected for a polymer of overall composition HOST:TBMA:STYR:PAG:Q = 50:30:6.7:10:3.3).

**Figure 8(g)-(i)** shows the analysis of the defined sequence polymer case with bound PAG and Q (composition **SA-BP-BQ-30-1.0**). **Figure 8(g)** reflects the homogeneous composition of the ensemble of chains. Once subjected to exposure and deprotection, the random packing of these identical chains and the random characteristics of exposure lead to a broad distribution in chain compositions in the final film. **Figure 8(h)** shows the resulting distribution for a diffusion length $d = 2$ nm. As found for the random copolymer cases, the MAA and MAA+HOST distributions are roughly Gaussian although with the defined sequence polymer their widths are somewhat narrower. **Figure 8(i)** shows histograms of the chain compositions after exposure/deprotection where $d = 5$ nm. As in the random copolymer case, there are a significant number of chains (fraction) where no MAA is formed, suggesting that acids are immediately quenched in the sequence A case (**Table 2**) even though PAGs and Qs are not adjacent on the chain, they are close enough to interact within the film (**Figure 7**). However, there is large number of chains (fraction) that have all TBMA groups in the volume converted to MAA,

creating a bimodal distribution. This means that at the assumed dose and diffusion length, when the acids formed can access a larger polymer film volume, TBMA is more likely to be fully deprotected. By starting with a uniform chain composition, rather than a very complex random composition distribution, and allowing each acid to react over a large volume, it is more likely that the deprotection process will go to completion. This bimodal form is replicated in the total fraction of ionizable groups per chain since each chain has an identical number of HOST groups. The MAA+HOST histogram is simply displaced to the right by a fixed amount equal to the fraction of HOST.

These distributions can be compared to a DSP with free PAG and Q (composition **SF-BP-BQ-30-1.0**), **Figures 8(j)-(l).** The ionizable fraction distributions in **Figures 8(b)** are narrower than in **Figure 8(k)** but their median values are very similar, indicating that it is the state of the PAG and Q, not the polymer structure, that most strongly influences solubility when the acid diffusion distance is 2 nm. At 5 nm, **Figure 8(l)** is intermediate between the distributions found in **Figures 8(c)** and **(i)**.

These findings raise the interesting question of how the combination of acid blur and defined sequences affect local solubility at line edges and hence line edge roughness, as well as what the resulting tradeoffs between resolution and line edge roughness might be. The present work was only performed for one dose, chosen to form 50% ionizable groups on average. These studies are currently being extended to examine a range of doses in order to evaluate the R-L-S tradeoff for DSPs.

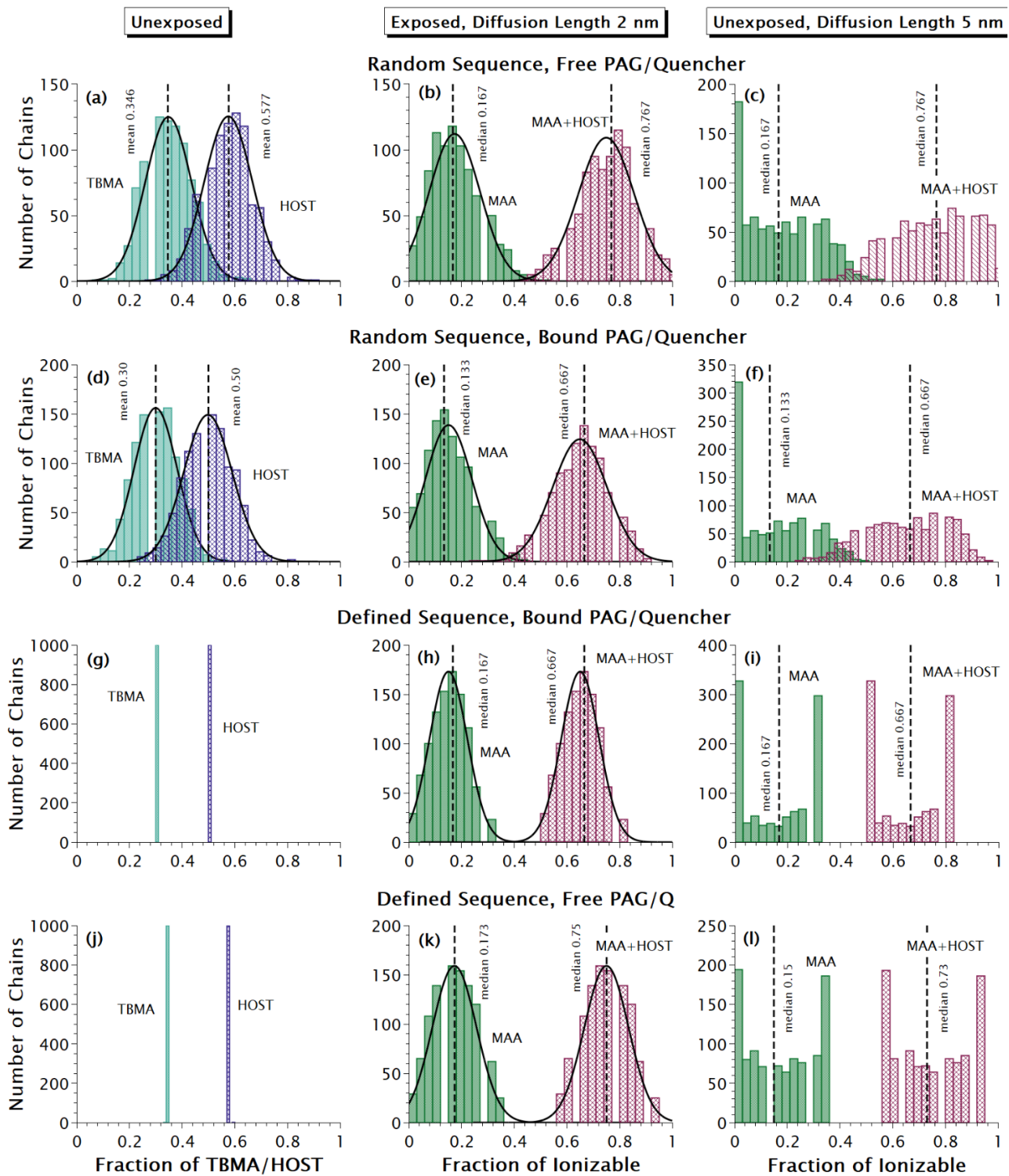

*Figure 8.* Numbers of polymer chains in the film slice having specific total monomers in them. All cases are for a chain length of 30 monomers and a polydispersity of 1.0. (a)-(c) are for a random sequence with free PAG and Q. (a) distributions of chains having specific fractions of TBMA and HOST in an unexposed film; (b) distributions of chains having specific fractions of MMA and HOST in an exposed and deprotected film, acid

*diffusion length of 2nm; (c) distributions of chains having specific fractions of MMA and HOST in an exposed and deprotected film, acid diffusion length of 5 nm. (d)-(f) are for a random sequence with bound PAG and Q. (d) distributions of chains having specific fractions of TBMA and HOST in an unexposed film; (e) distributions of chains having specific fractions of MMA and HOST in an exposed and deprotected film, acid diffusion length of 2nm; (f) distributions of chains having specific fractions of MMA and HOST in an exposed and deprotected film, acid diffusion length of 5 nm. (g)-(i) are for a defined sequence with bound PAG and Q. (g) distributions of chains having specific fractions of TBMA and HOST in an unexposed film; (h) distributions of chains having specific fractions of MMA and HOST in an exposed and deprotected film, acid diffusion length of 2nm; (i) distributions of chains having specific fractions of MMA and HOST in an exposed and deprotected film, acid diffusion length of 5 nm. (j)-(l) are for defined sequence with unbound PAG and Q. (j) distributions of chains having specific fractions of TBMA and HOST in an unexposed film; (k) distributions of chains having specific fractions of MMA and HOST in an exposed and deprotected film, acid diffusion length of 2nm; (l) distributions of chains having specific fractions of MMA and HOST in an exposed and deprotected film, acid diffusion length of 5 nm.*

**Conclusions**

In this paper, we report a computational study of the internal structure of polymer slices representing photoresists modeled on the ESCAP composition, containing TBMA, styrene and HOST on the polymer chains, and PAG and quencher either mixed in with the polymer chains, or bound to them. The polymer chains are either random or with a defined sequence, with variations in their length and polydispersity. Following assembly of the polymer slices, statistical analyses of the spatial correlations from PAG-PAG, TBMA-TBMA and PAG-Q have been made. Comparison of the internal structures of random and defined sequence polymers of the same compositions show differences in the nearest neighbor identities (within 1-2nm), but no differences beyond that distance that would indicate that DSPs have long-range order. These correlations are not affected by chain length or polydispersity. Imaging of these polymer slides using a simple exposure/deprotection model shows that the numbers of soluble chains formed using a 2 nm acid diffusion lengths have very similar distributions for both DSP and random copolymers with and without bound PAG and Q. Only one case was found to have distinctive solubility outcomes, DSP with either unbound or bound PAG and Quencher with an acid diffusion length of 5 nm. A remarkable bimodal distribution emerges in this case whose physical origin is not clearly related to spatial distributions of the resist components. How this finding depends on dose, and how it affects image quality are currently under study.

**Professional Biographies**

**William D. Hinsberg** received his doctorate in chemistry at the California Institute of Technology, followed by postdoctoral work at Stanford University. He joined IBM Corporation in 1982, working in thin-film process chemistry for magnetic recording devices and the chemistry of new photoresist materials. He has numerous awards, including an IBM Corporate Award for development of stochastic kinetics simulation methods. He has published more than 150 scientific papers and is inventor on more than thirty issued US patents. Currently, he is founder and president of Columbia Hill Technical Consulting.

**Frances A. Houle** received her doctorate in chemistry at the California Institute of Technology. She has held appointments at the IBM Research Division and Invisage Technologies, and is now a Senior Scientist at Lawrence Berkeley National Laboratory. Her current research focuses on chemical processes in EUV photolithography and solar photochemistry for energy conversion. She has received numerous awards including an IBM Corporate Award and the AVS John A. Thornton Memorial Award and Lecture. She has published more than 170 scientific papers, and has 28 issued U.S. Patents. ORCID: 0000-0001-5571-2548

**Disclosures**

The authors have no conflicts of interest to declare.

**Code, Data, and Materials Availability**

The data presented in this article are publicly available in Zenodo, DOI: 10.5281/zenodo.13765732. The archived version of the codes described in this manuscript can be freely accessed through the Zenodo link as well as GitHub (https://github.com/hnsbrg/PolyScope).

**Acknowledgments**

This paper is based upon work by Columbia Hill Technical Consulting (WDH), and by the Center for High Precision Patterning Science (CHiPPS), an Energy Frontier Research Center funded by the U.S. Department of Energy, Office of Science, Basic Energy Sciences (FAH). The authors are grateful to Dr. Matthias Mueller-Fischer for permission to use his algorithm from his PhD dissertation, and to Dr. Greg Wallraff for helpful discussions.

**References**

1. G. M. Gallatin, F. A. Houle and J. L. Cobb, "Statistical limitations of printing 50 and 80 nm contact holes by EUV lithography," *J Vac Sci Technol B* **21**(6), 3172-3176 (2003). https://doi.org/10.1116/1.1629294.
2. W. D. Hinsberg et al., "Contribution of EUV resist counting statistics to stochastic printing failures," *J Micro-Nanopattern* **20**(1), Artn 014603 (2021). https://doi.org/10.1117/1.Jmm.20.1.014603.
3. G. M. Gallatin, "Resist blur and line edge roughness," *Optical Microlithography XVIII, Pts 1-3* **5754**, 38-52 (2005). https://doi.org/10.1117/12.607233.
4. P. C. Tsiartas et al., "The mechanism of phenolic polymer dissolution: A new perspective," *Macromolecules* **30**(16), 4656-4664 (1997). https://doi.org/DOI 10.1021/ma9707594.
5. S. Fleischmann and V. Percec, "Synthesis of Well-Defined Photoresist Materials by SET-LRP," *J Polym Sci Pol Chem* **48**(10), 2251-2255 (2010). https://doi.org/10.1002/pola.24007.
6. Y. Guo et al., "Terpolymerization of Styrenic Photoresist Polymers: Effect of RAFT Polymerization on the Compositional Heterogeneity," *Macromolecules* **48**(11), 3438-3448 (2015). https://doi.org/10.1021/acs.macromol.5b00683.

7. G. G. Barclay et al., "Narrow polydispersity polymers for microlithography: Synthesis and properties," *Advances in Resist Technology and Processing Xiii* **2724**, 249-260 (1996). https://doi.org/Doi 10.1117/12.241823.
8. M. X. Wang et al., "Novel anionic photoacid generators (PAGs) and corresponding PAG bound polymers for sub-50 nm EUV lithography," *J Mater Chem* **17**(17), 1699-1706 (2007). https://doi.org/10.1039/b617133h.
9. D. N. Tuan, H. Yamamoto and S. Tagawa, "Study on Resist Performance of Polymer-Bound and Polymer-Blended Photo-Acid Generators," *Jpn J Appl Phys* **51**(8), Artn 086503 (2012). https://doi.org/10.1143/Jjap.51.086503.
10. K. Young-Gil et al., "A positive-working alkaline developable photoresist based on partially tert-Boc-protected calix[4]resorcinarene and a photoacid generator," *J Mater Chem* **12**(1), 53-57 (2002).
11. J. De Neve et al., "Sequence-definition from controlled polymerization: the next generation of materials," *Polym Chem-Uk* **9**(38), 4692-4705 (2018). https://doi.org/10.1039/c8py01190g.
12. Z. L. Li et al., "Hierarchical Nanomaterials Assembled from Peptoids and Other Sequence-Defined Synthetic Polymers," *Chem Rev* **121**(22), 14031-14087 (2021). https://doi.org/10.1021/acs.chemrev.1c00024.
13. F. Kaefer et al., "Controlled Sequence Photoresists from Polypeptoids," *J Photopolym Sci Tec* **35**(1), 29-33 (2022).
14. F. Käfer et al., "Polypeptoids: Exploring the Power of Sequence Control in a Photoresist for Extreme-Ultraviolet Lithography," *Adv Mater Technol-Us* **8**(23), ARTN 2301104 (2023). https://doi.org/10.1002/admt.202301104.
15. H. Ito et al., "Environmentally Stable Chemical Amplification Positive Resist; Principle, Chemistry, Contamination Resistance and Lithographic Feasibility," *J. Photopolym. Sci. Technol.* **7**, 433-448 (1994).
16. W. Hinsberg, "Polyscope," https://github.com/hnsbrg/PolyScope (2024).
17. M. Mueller, "The Structure of Dense Polymer Systems: Geometry, Algorithms, Software," p. 117, Eidgenossische Technische Hochschule, Zurich, Switzerland (1999).
18. G. M. Schmid et al., "Understanding molecular level effects during post exposure processing," *Advances in Resist Technology and Processing Xviii, Pts 1 and 2* **4345**, 1037-1047 (2001). https://doi.org/10.1117/12.436829.
19. G. P. Patsis, V. Constantoudis and E. Gogolides, "Effects of photoresist polymer molecular weight on line-edge roughness and its metrology probed with Monte Carlo simulations," *Microelectron Eng* **75**(3), 297-308 (2004). https://doi.org/10.1016/j.mee.2004.06.005.
20. G. P. Patsis et al., "Simulation of roughness in chemically amplified resists using percolation theory," *J Vac Sci Technol B* **17**(6), 3367-3370 (1999). https://doi.org/Doi 10.1116/1.591012.
21. R. C. Team, "R: A Language and Environment for Statistical Computing," R Foundation for Statistical Computing, Vienna, Austria (2023).
22. P. Team, "RStudio: Integrated Development Environment for R," Posit Software, PBS, Boston, MA (2024).
23. A. Baddeley, E. Rubak and R. Turner, *Spatial Point Patterns: Methodology and Applications*, R. Chapman and Hall/CRC Press, London (2015).
24. I. SRL, "QtiPlot," (2024).

25. P. de Schepper et al., "XAS Photoresists Electron/Quantum yields study with synchrotron light," *Advances in Patterning Materials and Processes Xxxii* **9425**, Artn 942507 (2015). https://doi.org/10.1117/12.2085951.
26. T. Kozawa and S. Tagawa, "Radiation Chemistry in Chemically Amplified Resists," *Jpn J Appl Phys* **49**(3), Artn 030001 (2010). https://doi.org/10.1143/Jjap.49.030001.
27. U. Okoroanyanwu, *Chemistry and Lithography, Second Edition, Vol. 2: Chemistry in Lithography, Ch.8*, SPIE (2023).
28. T. Watanabe et al., "CA resist with side chain PAG group for EUV resist," *J Photopolym Sci Tec* **19**(4), 521-524 (2006). https://doi.org/DOI 10.2494/photopolymer.19.521.
29. M. X. Wang et al., "Novel polymeric anionic photoacid generators (PAGs) and corresponding polymers for 193 nm lithography," *J Mater Chem* **16**(37), 3701-3707 (2006). https://doi.org/10.1039/b607918k.